\documentclass[a4paper,10pt]{article}
\usepackage{amsmath,amsfonts,amssymb,amsthm}
\usepackage{mathtools}
\usepackage{graphicx}
\usepackage{color}
\usepackage{amsmath}
\usepackage{comment}
\usepackage{a4wide}
\usepackage{url}
\usepackage{authblk}

\usepackage{bm}

\newcommand{\bx}{{\bm{x}}}

\newcommand{\CB}{\mathrm{B}}
\newcommand{\CW}{\mathrm{W}}
\newcommand{\CG}{\mathrm{G}}
\newcommand{\EB}{E_{\CB}}
\newcommand{\EG}{E_{\CG}}
\newcommand{\EW}{E_{\CW}}

\newcommand{\nd}{{\nu}}
\newcommand{\cnd}{{\nu'}}

\newcommand{\mw}{{\mathtt{\mu}}}

\newcommand{\vc}{{\tau}}

\newcommand{\bin}{\mathrm{bin}}

\newcommand{\card}[1]{\left\lvert #1 \right\rvert}

\def\multiset#1#2{\ensuremath{\left(\kern-.3em\left(\genfrac{}{}{0pt}{}{#1}{#2}\right)\kern-.3em\right)}}

\providecommand{\keywords}[1]
{ \medskip  
  \noindent
  \small	
  \textbf{\textit{Keywords---}} #1
}

\newtheorem{lemma}{Lemma}
\newtheorem{theorem}{Theorem}
\newtheorem{proposition}{Proposition}
\newtheorem{corollary}{Corollary}
\newtheorem{definition}{Definition}

\newtheorem{remark}{Remark}

\title{Winner Determination Algorithms \\ for Graph Games with Matching Structures\footnote{A preliminary version of this paper appears in Proceedings of the 33rd International Workshop on Combinatorial Algorithms (IWOCA 2022),  Lecture Notes in Computer Science, Vol. 13270, pp. 509--522, Springer, 2022~\cite{Yoshiwatari2022}.}}
\author[1]{Tesshu Hanaka\thanks{\texttt{hanaka@inf.kyushu-u.ac.jp}}}
\affil[1]{Department of Informatics, Kyushu University, Fukuoka, Japan}

\author[2]{Hironori Kiya\thanks{\texttt{h-kiya@econ.kyushu-u.ac.jp}}}
\affil[2]{Department of Economic Engineering, Kyushu University, Fukuoka, Japan}

\author[3]{Hirotaka Ono\thanks{\texttt{ono@nagoya-u.jp}}}
\affil[3]{Department of Mathematical Informatics, Nagoya University, Aichi, Japan}

\author[3]{Kanae Yoshiwatari\thanks{\texttt{yoshiwatari.kanae.w1@s.mail.nagoya-u.ac.jp}}}

\date{}

\begin{document}

\maketitle

\begin{abstract}
Cram, Domineering, and Arc Kayles are well-studied combinatorial games. 
They are interpreted as edge-selecting-type games on graphs, and the selected edges during a game form a matching. 
In this paper, we define a generalized game called {Colored Arc Kayles}, which includes these games. 
{Colored Arc Kayles} is played on a graph whose edges are colored in black, white, or gray, and black (resp., white) edges can be selected only by the black (resp., white) player, although gray edges can be selected by both black and white players.  
We first observe that the winner determination for {Colored Arc Kayles} can be done in $O^*(2^n)$ time by a simple algorithm, where $n$ is the order of a graph. 
We then focus on the vertex cover number, which is linearly related to the number of turns,   
and show that Colored Arc Kayles, BW-Arc Kayles, and Arc Kayles are solved in time $O^*(1.4143^{\vc^2+3.17\vc})$, $O^*(1.3161^{\vc^2+4{\vc}})$, and
$O^*(1.1893^{\vc^2+6.34{\vc}})$, respectively, where $\vc$ is the vertex cover number. 
Furthermore, we present an $O^*((n/\nd+1)^{\nd})$-time algorithm for Arc Kayles, where $\nd$ is neighborhood diversity.  
We finally show that Arc Kayles on trees can be solved in $O^* (2^{n/2})(=O(1.4143^n))$ time, which improves $O^*(3^{n/3})(=O(1.4423^n))$ by a direct adjustment of the analysis of Bodlaender et al.'s $O^*(3^{n/3})$-time algorithm for Node Kayles. 

\keywords{Arc Kayles \and Combinatorial Game Theory \and Exact Exponential-Time Algorithm \and Vertex Cover \and Neighborhood Diversity.}
\end{abstract}

\section{Introduction}
\subsection{Background and Motivation} 
Cram, Domineering, and Arc Kayles are well-studied two-player mathematical games and 
interpreted as combinatorial games on graphs. 
Domineering (also called Stop-Gate) was   
introduced by G\"oran Andersson around 1973 under the name of Crosscram~\cite{conway2000numbers,gardner1974mathematical}.
Domineering is usually played on a checkerboard. The two players are denoted by
Vertical and Horizontal. Vertical (resp., Horizontal) player is
only allowed to place its dominoes vertically (resp., horizontally) on the board.
Note that placed dominoes are not allowed to overlap. 
If no place is left to place a domino, the player in the turn loses the game. Domineering is a partisan game, where players use different pieces. The impartial version of the game is Cram, where two players can place dominoes both vertically and horizontally. 

\begin{figure}[tbp]
 \centering
 \includegraphics[width=0.8\linewidth]{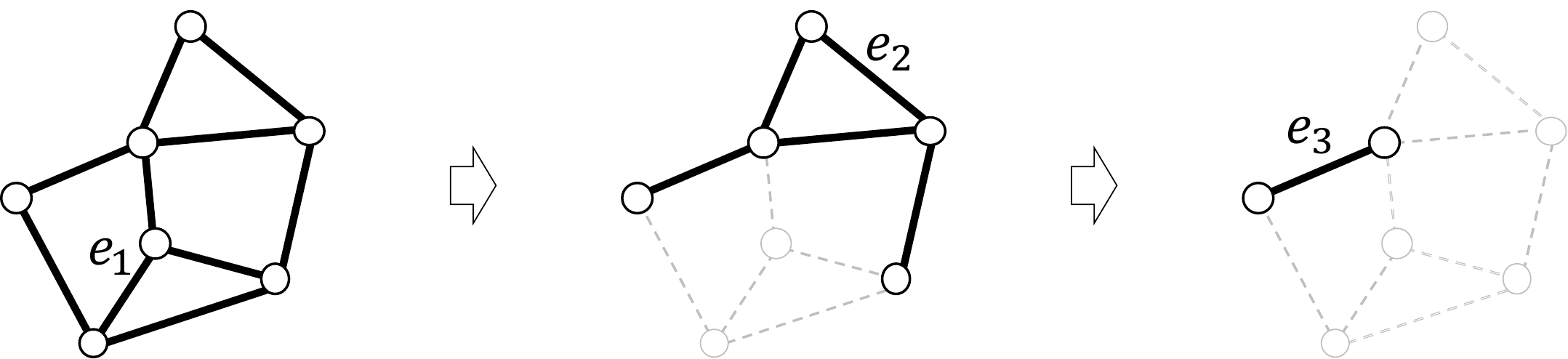}
 \caption{A play example of Arc Kayles}
 \label{ex:arcKayles}
\end{figure}

An analogous game played on an undirected graph $G$ is Arc Kayles.
In Arc Kayles, the action of a player in a turn is to select an edge of $G$, and then the selected edge and its neighboring edges are removed from $G$. If no edge remains in the resulting graph, the player in the turn loses the game.  
Figure \ref{ex:arcKayles} is a play example of Arc Kayles. In this example, the first player selects edge $e_1$, and then the second player selects edge $e_2$. By the first player selecting edge $e_3$, no edge is left; the second player loses.  
Note that the edges selected throughout a play form a maximal matching on the graph.

Similarly, we can define BW-Arc Kayles, which is played on an undirected graph with black and white edges. The rule is the same as the ordinary Arc Kayles except that the black (resp., white) player can select only black (resp., white) edges. 
Note that Cram and Domineering are respectively interpreted as Arc Kayles and BW-Arc Kayles on a two-dimensional grid graph, which is the graph Cartesian product of two path graphs.  

To focus on the common nature of such games with matching structures, we newly define {Colored Arc Kayles}. {Colored Arc Kayles} is played on a graph whose edges are colored in black, white, or gray, and black (resp., white) edges can be selected only by the black (resp., white) player, though grey edges can be selected by both black and white players. 
{BW-Arc Kayles} and ordinary {Arc Kayles} are special cases of {Colored Arc Kayles}. In this paper, we investigate {Colored Arc Kayles} from the algorithmic point of view.

\subsection{Related work}
\subsubsection{Cram and Domineering}
Cram and Domineering are well studied in the field of combinatorial game theory. In \cite{gardner1974mathematical}, Gardner gives winning strategies for some simple cases. For Cram on $a\times b$ board, the second player can always win if both $a$ and $b$ are even, and the first player can always win if one of $a$ and $b$ is even and the other is odd. This can be easily shown by the so-called Tweedledum and Tweedledee strategy. 
For specific sizes of boards, computational studies have been conducted~\cite{uiterwijk2019solving}. 
In \cite{uiterwijk2018construction}, Cram's endgame databases for all board sizes with at most 30 squares are constructed. 
As far as the authors know, the complexity to determine the winner for Cram on general boards still remains open.

Finding the winning strategies of Domineering for specific sizes of boards by using computer programs is well studied. 
For example, the cases of $8\times 8$ and $10\times 10$ are solved in 2000~\cite{BREUKER2000195} and 2002~\cite{bullock2002domineering}, respectively. The first player wins in both cases. Currently, the status of boards up to $11\times 11$ is known~\cite{uiterwijk201611}. 
In \cite{uiterwijk2015new}, endgame databases for all single-component positions up to 15 squares for Domineering are constructed.
The complexity of Domineering on general boards also remains open. Lachmann, Moore, and Rapaport show that the winner and a winning strategy Domineering on $m\times n$ board can be computed in polynomial time for $m\in \{1, 2, 3, 4, 5, 7, 9, 11\}$ and all $n$~\cite{lachmann2000wins}.

\subsubsection{Kayles, Node Kayles, and Arc Kayles}
Kayles is a simple impartial game, introduced by Henry Dudeney in 1908~\cite{dudeney2002canterbury}.  
The name ``Kayles'' derives from French word ``quilles'', meaning ``bowling''. The rule of Kayles is as follows.  
Given bowling pins equally spaced in a line, players take turns to knock out either one pin or two adjacent pins, until all the pins are gone.
As graph generalizations, Node Kayles and Arc Kayles are introduced by Schaefer~\cite{SCHAEFER1978185}. 
Node Kayles is the vertex version of Arc Kayles. Namely, 
the action of a player is to select a vertex instead of an edge, and then the selected vertex and its neighboring vertices are removed. 
Note that both generalizations can describe the original Kayles; Kayles is represented as Node Kayles on sequentially linked triangles or as Arc Kayles on a caterpillar graph.

Node Kayles is known to be PSPACE-complete \cite{SCHAEFER1978185}, whereas the winner determination is solvable in polynomial time on graphs of bounded asteroidal numbers such as cocomparability graphs and cographs by using Sprague-Grundy theory~\cite{BODLAENDER2002}. 
For general graphs, Bodlaender et al. propose an $O(1.6031^n)$-time algorithm~\cite{BODLAENDER2015165}. Furthermore, they show that the winner of Node Kayles can be determined in time $O(1.4423^{n})$ on trees. In \cite{Kobayashi2018}, Kobayashi sophisticates the analysis of the algorithm in \cite{BODLAENDER2015165} from the perspective of the parameterized complexity and shows that it can be solved in time $O^*(1.6031^{\mw})$, where $\mw$ is the modular width of an input graph\footnote[1]{The $O^*(\cdot)$ notation suppresses polynomial factors in the input size.}. He also gives an $O^*(3^{\vc})$-time algorithm, where $\vc$ is the vertex cover number, 
and a linear kernel when parameterized by neighborhood diversity.

Different from Node Kayles, the complexity of Arc Kayles has remained open for more than 30 years. Even for subclasses of trees, not much is known. For example, Huggans and Stevens study Arc-Kayles on subdivided stars with three paths~\cite{huggan2016polynomial}. 
To our best knowledge, 
no exponential-time algorithm for Arc Kayles is presented except for an $O^*(4^{\vc^2})$-time algorithm proposed in \cite{LM2014}.

\subsection{Our contribution}

In this paper, we address winner determination algorithms for Colored Arc Kayles. 
We first propose an $O^*(2^n)$-time algorithm for Colored Arc Kayles. 
Note that this is generally faster than applying the Node Kayles algorithm  to the line graph of an instance of Arc Kayles;  
it takes time $O(1.6031^m)$, where $m$ is the number of the original edges.
We then focus on algorithms based on graph parameters.
We present an $O^*(1.4143^{\vc^2+3.17\vc})$-time algorithm for Colored Arc Kayles, where $\vc$ is the vertex cover number. 
The algorithm runs in time $O^*(1.3161^{\vc^2+4\vc})$ and $O^*(1.1893^{\vc^2+6.34\vc})$ for BW-Arc Kayles, and Arc Kayles, respectively. 
This is faster than the previously known time complexity $O^*(4^{\vc^2})$ in  \cite{LM2014}. 

On the other hand, we give a bad instance for the proposed algorithm, which implies the running time analysis is asymptotically tight.
Furthermore, we show that the winner of Arc Kayles can be determined in time $O^*((n/\nd+1)^{\nd})$,  where $\nd$ is the neighborhood diversity of an input graph. This analysis is also asymptotically tight, because there is an instance having $(n/\nd-o(1))^{\nd(1-o(1))}$.
We finally show that the winner determination of Arc Kayles on trees can be solved in $O^*(2^{n/2})=O(1.4143^n)$ time, which improves $O^*(3^{n/3})(=O(1.4423^n))$ by a direct adjustment of the analysis of Bodlaender et al.'s $O^*(3^{n/3})$-time algorithm for Node Kayles. 

\section{Preliminaries}\label{sec:preliminaries}
\subsection{Notations and terminology}
Let $G=(V,E)$ be an undirected graph. We denote $n=\card{V}$ and $m=\card{E}$, respectively. 
For an edge $e=\{u,v\}\in E$, we define $\Gamma(e)=\{e'\mid e\cap e'\neq \emptyset\}$.
For a graph $G=(V,E)$ and a vertex subset $V'\subseteq V$, we denote by $G[V']$ the subgraph induced by $V'$. For simplicity, we denote $G-v$ instead of $G[V\setminus \{v\}]$. For an edge subset $E'$, we also denote by $G-E'$ the subgraph obtained from $G$ by removing all edges in $E'$ from $G$. 
A vertex set $S$ is called a \emph{vertex cover} if  $e\cap S\neq \emptyset$ for every edge $e\in E$. We denote by $\vc$ the size of a minimum vertex cover of $G$.
Two vertices $u,v\in V$ are called \emph{twins} if $N(u)\setminus \{v\}=N(v)\setminus \{u\}$.

\begin{definition}\label{def:nd}
The \emph{neighborhood diversity} $\nd(G)$ of $G=(V,E)$ is defined as the minimum number $w$ such that $V$ can be partitioned into $w$ vertex sets of twins. 
\end{definition}

In the following, we simply write $\nd$ instead of $\nd(G)$ if no confusion arises. We can compute the neighborhood diversity of $G$ and the corresponding partition in polynomial time \cite{Lampis2012}. For any graph $G$,  $\nd \le 2^{\vc}+\vc$ holds.

\subsection{Colored Arc Kayles}
Colored Arc Kayles is played on a graph $G=(V,\EG\cup \EB\cup \EW)$, where $\EG,\EB, \EW$ are mutually disjoint. The subscripts $\CG$, $\CB$, and $\CW$ of $\EG, \EB, \EW$ respectively, stand for gray, black, and white. 
For every edge $e\in \EG \cup \EB \cup \EW$, let $c(e)$ be the color of $e$, that is, $c(e)=\CG$ if $e\in \EG$, $\CB$ if $e\in \EB$, and $\CW$ if $e\in \EW$.
If $\{u,v\}\not\in \EG \cup \EB \cup \EW$, we set $c(\{u,v\})=\emptyset$ 
for convenience. 
As explained below, the first (black or $\CB$) player can choose only gray or black edges, and the second (white or $\CW$) player can choose only gray or white edges. 

Two players alternatively choose an edge of $G$. Player B can choose an edge in $\EG\cup \EB$ and player W can choose an edge in $\EG\cup \EW$.
That is, there are three types of edges; $\EB$ is the set of edges that only the first player can choose, $\EW$ is the set of edges that only the second player can choose, and $\EG$ is the set of edges that both the first and second players can choose. Once an edge $e$ is selected, the edge and its neighboring edges (i.e., $\Gamma(e)$) are removed from the graph, and the next player chooses an edge of $G- \Gamma(e)$.  
The player that can take no edge loses the game. 
Since (Colored) Arc Kayles is a two-person zero-sum perfect information game and ties are impossible, one of the players always has a winning strategy. 
We call the player having a winning strategy the \emph{definite winner}, or simply \emph{winner}. 

The problem that we consider in this paper is defined as follows: 
\begin{description}
\item {\bf Input}: $G=(V,\EG\cup \EB\cup \EW)$, active player in $\{\CB,\CW\}$. 
\item {\bf Question}: Suppose that players $\CB$ and $\CW$ play Colored Arc Kayles on $G$ from the active player's turn. Which player is the winner?
\end{description}
Remark that if $\EB=\EW=\emptyset$, Colored Arc Kayles is equivalent to Arc Kayles and if $\EG=\emptyset$, it is equivalent to BW-Arc Kayles.

To simply represent the definite winner of Colored Arc Kayles, we introduce two Boolean functions $f_{\CB}$ and $f_{\CW}$. The $f_{\CB}(G)$ is defined such that $f_{\CB}(G)=1$ if and only if the winner of Colored Arc Kayles on $G$ from player B's turn is player B. Similarly, $f_{\CW}(G)$ is the function such that $f_{\CW}(G)=1$ if and only if the winner of Colored Arc Kayles on $G$ from player W's turn is the player W. If two graphs $G$ and $G'$ satisfy that 
$f_{\CB}(G)=f_{\CB}(G')$ and $f_{\CW}(G)=f_{\CW}(G')$, we say that \emph{$G$ and $G'$ have the same game value on Colored Arc Kayles}.

\section{
Basic Algorithm }\label{sec:Exalgo}
In this section, we show that the winner of \emph{Colored} Arc Kayles on $G$ can be determined in time $O^*(2^n)$. We first observe that  the following lemma holds by the definition of the game.
\begin{lemma}\label{lem:winner:Colored_Arc_Kayles}
Suppose that Colored Arc Kayles is played on $G=(V,\EG\cup \EW\cup \EB)$. 
Then, player $\CB$ (resp., $\CW$) wins on $G$ with player $\CB$'s (resp., $\CW$'s) turn if and only if there is an edge $\{u,v\}\in \EG\cup \EB$ (resp., $\{u,v\}\in \EG\cup \EW$) such that player $\CW$ (resp., $\CB$) loses on $G-u-v$ with player $\CW$'s (resp., $\CB$'s) turn. 
\end{lemma}
This lemma is interpreted by the following two recursive formulas:
\begin{align}\label{eq:recursion1}
    f_{\CB}(G) & = \bigvee_{\{u,v\}\in{\EG\cup \EB}}\lnot \left(f_{\CW}(G-u-v)\right),\\ \label{eq:recursion2} 
    f_{\CW}(G) & = \bigvee_{\{u,v\}\in{\EG\cup \EW}}\lnot \left(f_{\CB}(G-u-v)\right).      
\end{align}
By these formulas, we can determine the winner of $G$ with either first or second player's turn by computing $f_{\CB}(G)$ and $f_{\CW}(G)$ for all induced subgraphs of $G$. Since the number of all induced subgraphs of $G$ is $2^n$,  it can be done in time $O^*(2^n)$ by a standard dynamic programming algorithm.

\begin{theorem}\label{thm:Colored_Arc_Kayles:Exp}
The winner of Colored Arc Kayles can be determined in time $O^*(2^n)$.
\end{theorem}

\section{FPT algorithm parameterized by vertex cover}\label{sec:vc}
In this section, we propose winner determination algorithms for Colored Arc Kayles parameterized by the vertex cover number. As mentioned in Introduction, the selected edges in a play of Colored Arc Kayles form a matching. This implies that the number of turns is bounded above by the maximum matching size of $G$ and thus by the vertex cover number.  
Furthermore, the vertex cover number of the input graph is bounded by twice of the number of turns of Arc Kayles. 
Intuitively, we may consider that a game taking longer turns is harder to analyze than games taking shorter turns. In that sense, the parameterization by the vertex cover number is quite natural. 

In this section, we propose an $O^*(1.4143^{\vc^2+3.17\vc})$-time algorithm for Colored Arc Kayles, where $\vc$ is the vertex cover number of the input graph.
It utilizes similar recursive relations shown in the previous section, 
but we avoid to enumerate all possible positions by utilizing equivalence classification. 

Before explaining the equivalence classification, we give a simple observation based on isomorphism. The isomorphism on edge-colored graphs is defined as follows. 

\begin{definition}\label{def:isomorphic:colored}
Let $G^{(1)}=(V^{(1)},\EG^{(1)}\cup \EB^{(1)}\cup \EW^{(1)})$ and $G^{(2)}=(V^{(2)},\EG^{(2)}\cup \EB^{(2)}\cup \EW^{(2)})$ be edge-colored graphs. 
Then 
$G^{(1)}$ and $G^{(2)}$ are called \emph{isomorphic} if for any pair of $u,v\in V$ there is a bijection $f: V^{(1)}\to V^{(2)}$ such that (i) $\{u,v\}\in \EG^{(1)}$ if and only if $\{f(u),f(v)\}\in \EG^{(2)}$, (ii) $\{u,v\}\in \EB^{(1)}$ if and only if $\{f(u),f(v)\}\in \EB^{(2)}$, and (iii) $\{u,v\}\in \EW^{(1)}$ if and only if $\{f(u),f(v)\}\in \EW^{(2)}$. 
\end{definition}

The following proposition is obvious. 

\begin{proposition}\label{prop:isomorphic}
If edge-colored graphs $G^{(1)}$ and $G^{(2)}$ are isomorphic, 
$G^{(1)}$ and $G^{(2)}$ have the same game value for Colored Arc Kayles.
\end{proposition}

Let $S$ be a vertex cover of $G=(V,\EG\cup \EW\cup \EB)$, that is, any $e=\{u,v\}\in \EG \cup \EW \cup \EB$ satisfies that $\{u,v\}\cap S\neq \emptyset$. 
Note that for $v\in V\setminus S$, $N(v)\subseteq S$ holds. 
We say that two vertices $v,v'\in V\setminus S$ are \emph{equivalent with respect to $S$ in $G$} if $N(v)=N(v')$ and $c(\{u,v\})=c(\{u,v'\})$ holds for $\forall u\in N(v)$. If two vertices $v,v'\in V\setminus S$ are equivalent with respect to $S$ in $G$, $G-u-v$ and $G-u-v'$ are isomorphic because the bijective function swapping only $v$ and $v'$ satisfies the isomorphic condition. Thus, we have the following lemma.
\begin{lemma}\label{lem:isomorphic:type}
Suppose that two vertices $v,v'\in V\setminus S$ are equivalent with respect to $S$ in $G$. Then, for any $u\in N(v)$, $G-u-v$ and $G-u-v'$ have the same game value.
\end{lemma}

By the equivalence with respect to $S$, we can split $V\setminus S$ into equivalence classes. Note here that the number of equivalence classes is at most $4^{\card{S}}$, because for each $u\in S$ and $v \in V\setminus S$, edge $\{u,v\}$ does not exist, or it can be colored with one of three colors if exists; 
we can identify an equivalent class with $\bx\in \{\emptyset,\CG,\CB,\CW\}^{S}$, a 4-ary vector with length $\card{S}$. For $S'\subseteq S$, let $\bx[S']$ denotes
the vector by dropping the components of $\bx$ except the ones corresponding to $S'$. Also for $u\in S$, $\bx[u]$ denotes the component  corresponding to $u$ in $\bx$. 
Then, $V$ is partitioned into $V_S^{(\bx)}$'s, where $V_S^{(\bx)}=\{v\in V\setminus S \mid \forall u\in S: c(\{v,u\})=\bx[u]\}$. We arbitrarily define  the representative of non-empty $V_S^{(\bx)}$ (e.g., the vertex with the smallest ID), which is denoted by $\rho(V_S^{(\bx)})$. By using $\rho$, we also define the representative edge set by
\begin{align*}
    E^R(S)=\bigcup_{\bx\in \{\emptyset,\CG,\CB,\CW\}^{S}} \{\{u,\rho(V_S^{(\bx)})\} \in \EG\cup \EB \cup \EW \mid u \in S\}.
\end{align*}
By Lemma \ref{lem:isomorphic:type}, we can assume that both players choose an edge only in $E^R(S)$, which enables to modify the recursive equations (\ref{eq:recursion1}) and (\ref{eq:recursion2}) as follows: 
For a vertex cover $S$ of $G$, we have  
\begin{align}\label{eq:recursion3}
    f_B(G) & = \bigvee_{\{u,v\}\in{(\EG\cup \EB)\cap (E^R(S) \cup S\times S)}}\lnot \left(f_W(G-u-v))\right),\\ \label{eq:recursion4} 
    f_W(G) & = \bigvee_{\{u,v\}\in{(\EG\cup \EW)\cap (E^R(S) \cup S\times S)}}\lnot \left(f_B(G-u-v))\right).      
\end{align}
Note that this recursive formulas imply that the winner of Colored Arc Kayles can be determined in time $O^*((\tau^2 +\tau\cdot 4^{\tau})^{\tau})=
O^*((4^{\tau+\log_4 \tau})^{\tau})=O^*(4^{\tau^2+\tau\log_4 \tau})$ $=O^*(5.6569^{\tau^2})$, because the recursions are called at most $\card{S}$ times and $\tau+\log_4 \tau \le 1.25 \tau$ for $\tau\ge 1$. 

\medskip 

In the following, we give a better estimation of the number of induced subgraphs appearing in the recursion. Once such subgraphs are listed up, we can apply a standard dynamic programming to decide the necessary function values, or we can compute $f_B$ and $f_W$ according to the recursive formulas with memorization, by which we can skip redundant recursive calls. 
In order to estimate the number of induced subgraphs appearing in the recursion, we focus on the fact that the position of a play in progress corresponds to the subgraph induced by a matching. 
 \begin{lemma}\label{lem:positions:vc}
The number of nodes in recursion trees of equations (\ref{eq:recursion3}) and (\ref{eq:recursion4}) for Colored Arc Kayles is $O((r+1)^{\card{S}^2/4}3^{\card{S}}\card{S}^2)$, where $r$ is the used colors.
 \end{lemma}
\begin{proof}
Suppose that $S$ is a vertex cover of $G$ and players play Colored Arc Kayles on $G$. At some point, some edges selected by players together with their neighboring edges are removed, and the left subgraph represents a game position. Note that at least one endpoint of such a selected edge is in $S$, and selected edges form a matching. 
To define such a subgraph, let us imagine that some matching $M$ is the set of edges that have been selected until the point. 
Although we do not specify $M$, the $M$ defines a partition $(X,Y,Z)$ of $S$; 
$X=\{u\in S \mid  \exists \{u,v\}\in M: v\in S\}$,
$Y=\{u\in S \mid  \exists \{u,v\}\in M: v\not\in S\}$,
and $Z=S\setminus (X\cup Y)$. 
We now count the number of positions having a common $(X,Y,Z)$. 
Since $X$ and $Y$ are removed, the remaining vertices in $V\setminus S$ are classified into $V_Z^{(\bx)}$'s 
$\bx \in \{\emptyset,\CG,\CB,\CW\}^{\card{Z}}$.   
This is the common structure defined by $(X,Y,Z)$, and the positions vary as $\card{Y}$ vertices in $\bigcup_{\bx \in \{\emptyset,\CG,\CB,\CW\}^{\card{Z}}} V_Z^{(\bx)}$ 
are matched with $Y$. Thus, we estimate the number of positions by counting the number of choices of $\card{Y}$ vertices in $\bigcup_{\bx \in \{\emptyset,\CG,\CB,\CW\}^{\card{Z}}} V_Z^{(\bx)}$. Here, let $\gamma$ be the number of used colors of edges. For example, BW-Arc Kayles and ordinary Arc Kayles use $\gamma=2$ and $\gamma=1$ colors, respectively, and which may reduce the numbers of $\bx$'s for smaller $\gamma$. Then, it is above bounded
by the multiset coefficient of $(\gamma+1)^{\card{Z}}$ multichoose $\card{Y}$, 
\[ 
\multiset{(\gamma+1)^{\card{Z}}}{\card{Y}}=\binom{(\gamma+1)^{\card{Z}}+\card{Y}-1}{\card{Y}}\le (\gamma+1)^{\card{Z}\card{Y}}. 
\]
This is an upper bound of the number of subgraphs to consider with respect to $(X,Y,Z)$. By considering all possible $(X,Y,Z)$, the total number of subgraphs is bounded by 
\begin{align*}
\sum_{X,Y,Z\subseteq S}(\gamma+1)^{\card{Z}\card{Y}}      
& \le 3^{\card{S}}\cdot {\card{S}}\sum_{y=0}^{\card{S}}(\gamma+1)^{y({\card{S}}-y)}\le 3^{\card{S}} \cdot {\card{S}}^2 \cdot {(\gamma+1)^{\frac{{\card{S}}^2}{4}}}, 
\end{align*}
where the first inequality comes from the choices of $X$, $Y$, and $Z$ from $S$. 
\end{proof}

The following theorem immediately holds by Lemma \ref{lem:positions:vc} and the fact that a minimum vertex cover of $G$ can be found in time $O^*(1.2738^\vc)$, where $\vc$ is the vertex cover number of $G$~\cite{chen2010improved}. 
\begin{theorem}\label{thm:algo:vc}
The winners of Colored Arc Kayles, BW-Arc Kayles, and Arc Kayles can be determined in time $O^*(1.4143^{\vc^2+3.17\vc})$, 
$O^*(1.3161^{\vc^2+4\vc})$, and  $O^*(1.1893^{{\vc}^2+6.34{\vc}})$, respectively, where $\vc$ is the vertex cover number of a graph.  
\end{theorem}

\begin{figure}[htbp]
 \centering
 \includegraphics[width=0.8\linewidth]{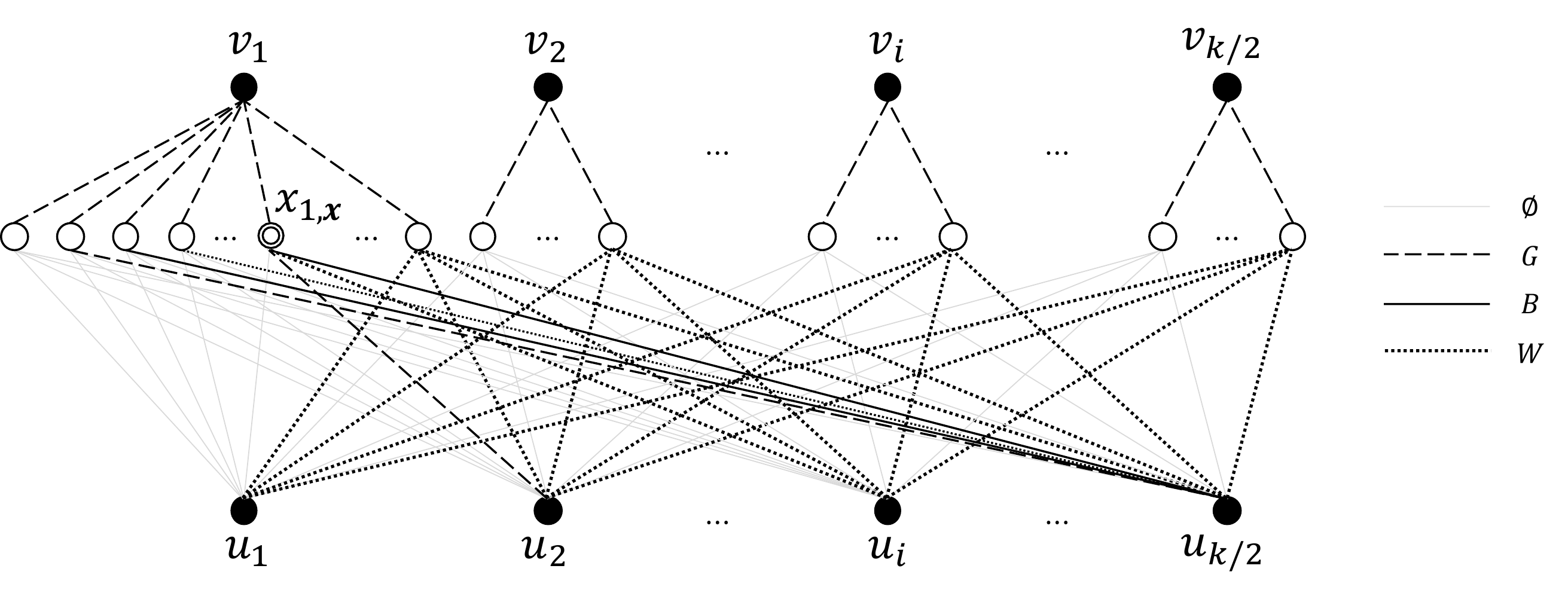}
 \caption{The constructed graph $G=(U\cup V\cup X, E)$.}
 \label{fig:lower}
\end{figure}

We have shown that the winner of Arc Kayles can be determined in time $O^*(1.1893^{{\vc}^2+6.34{\vc}})$. 
The following theorem shows that the analysis is asymptotically tight,  
which implies that for further improvement, we need additional techniques apart from ignoring vertex-cover-based isomorphic positions. We here give such an example in Figure \ref{fig:lower}. 

\begin{theorem}\label{thm:lower:vc}
There is a graph for which the algorithm requires $2^{\vc^2/2}$ recursive calls for Colored Arc Kayles.
\end{theorem}

\begin{proof}
In this proof, we assume that $k$ is a multiple of $4$. 
We explain how we systematically construct such a graph $G$ (see Figure \ref{fig:lower}). 
Let $k$ be an even number. 
We first define $U=\{u_1,\ldots, u_{k/2}\}$ and $V= \{v_1, \ldots, v_{k/2}\}$ as vertex sets. The union $U\cup V$ will form a vertex cover after the graph $G$ are constructed.
For every 4-ary vector $\bx\in \{\emptyset, \CG,\CB,\CW\}^{U}$ and every $i\in \{1,\ldots,k/2\}$, 
we define $x_{i,\bx}$ as a vertex, and let $X$ be the collections of $x_{i,\bx}$'s, i.e., $X=\{x_{i,\bx} \mid \bx\in \{\emptyset, \CG,\CB,\CW\}^{U}, i\in \{1,\ldots,k/2\}\}$. These are the vertices of $G$. 
We next define the set of edges of $G$. 
We connect $v_i$ and $x_{i,\bx}$'s by black edges and by white edges for each $i\in \{1,2,\ldots,k/4\}$ and $\{k/4+1,k/4+2,\ldots,k/2\}$, respectively.  
Furthermore, we connect $x_{i,\bx}$ 
and $u\in U$ 
for each $\bx$ so that $\{u,x_{i,\bx}\}$ has 
color $\bx[u]$; if $\bx[u]=\emptyset$, no edge exists between $x_{i,\bx}$ and $u\in U$. Figure~\ref{fig:lower} shows an example how we connect $x_{1,\bx}$ and $u_i$'s, where 
$\bx[u_1]=\emptyset$, $\bx[u_2]=\CG$, $\bx[u_i]=\CW$, and $\bx[u_{k/2}]=\CB$. Notice that in Figure \ref{fig:lower}, $\{x_{1,\bx},u_1\}$ is connected with edge $\emptyset$ for explanation, which means that there is no edge between $x_{1,\bx}$ and $u_1$. 
Note that the number of vertices in $G$ is $\card{U}+\card{V}+\card{X}=k+4^{k/2}k/2$. 
Moreover, 
$U$, $V$, and $X$ form independent sets and $X$ separates $U$ and $V$.
Thus, $U\cup V$ forms a vertex cover of size $k$ in $G$.

We are ready to explain that $G$ has different $2^{\vc^2/2}$ subgraphs called by the algorithm. 
Starting from $G$, we call the recursive formulas (\ref{eq:recursion3}) and (\ref{eq:recursion4}) $k/2$ times by selecting edges incident to only $v_i$'s. 
Then, all the vertices in $V$ are removed from $G$, and the neighbors of remaining $x_{i,\bx}$'s are in $U$. That is, each $x_{i,\bx}$ has its inherent set of neighbors, and thus 
$x_{i,\bx}$'s are not equivalent each other for vertex cover $U\cup V$.
This implies that if the set of removed edges are different, the resulting subgraphs are also different. 

In a step before $k/2+1$, an edge connecting some $v_i$ is removed, and such an edge is chosen from $\{\{v_i,x_{i,\bx}\}\mid \bx\in \{\emptyset, \CG,\CB,\CW\}^{U}\}$, that is, the number of candidates is $4^{k/2}$ for each $i$. 
Thus, the total way to choose edges is $(4^{k/2})^{k/2}=4^{k^2/4}=2^{k^2/2}$; at least $2^{k^2/2}$ recursion calls occur.  
\end{proof}

By the similar construction, we can show the following theorem.

\begin{theorem}\label{thm:lower:vc2}
There is a graph for which the algorithm requires $1.3161^{\vc^2}$ and $1.1893^{\vc^2}$ recursive calls for BW-Arc Kayles and Arc Kayles, respectively.
\end{theorem}

\begin{remark}
Although Theorems \ref{thm:lower:vc} and \ref{thm:lower:vc2} give lower bounds on the running time of the vertex cover-based algorithms, the proof implies a stronger result. In the proof of Theorem \ref{thm:lower:vc}, we use ID's of the vertices in $U$. By connecting $i$ pendant vertices to $u_i$, we can regard them as ID of $u_i$. Furthermore, we make $U$ a clique by adding edges. These make the graphs not automorphic, which implies that the time complexity of an algorithm utilizing only isomorphism is at least the value shown in Theorems \ref{thm:lower:vc} or \ref{thm:lower:vc2}.    \end{remark}

\section{XP algorithm parameterized by neighborhood diversity}\label{sec:nd}
In this section, we deal with neighborhood diversity $\nd$, which is a more general parameter than vertex cover number. We first give an $O^*((n/\nd)^{\nd})$-time algorithm for Arc Kayles. This is an XP algorithm parameterized by neighborhood diversity. On the other hand, we show that there is a graph having at least $O^*((n/\nd)^{{\rm \Omega}(\nd)})$ non-isomorphic induced subgraphs, which implies the analysis of the proposed algorithm is 
asymptotically tight.

By Proposition \ref{prop:isomorphic}, if we list up all non-isomorphic induced subgraphs, the winner of Arc Kayles can be determined by using recursive formulas (\ref{eq:recursion1}) and (\ref{eq:recursion2}).
Let $\mathcal{M}=\{M_1,M_2,\ldots,M_{\nd}\}$ be a partition such that $\bigcup_i M_i=V$ and vertices of $M_i$ are twins each other.  We call each $M_i$ a \emph{module}.
We can see that non-isomorphic induced subgraphs of $G$ are identified by how many vertices are selected from which module.

\begin{lemma}\label{lem:positions:nd}
The number of non-isomorphic induced subgraphs of a graph of neighborhood diversity $\nd$ is at most $(n/\nd+1)^{\nd}$.
\end{lemma}
\begin{proof}
By the definition of neighborhood diversity, vertices in a module are twins each other. Therefore, the number of non-isomorphic induced subgraphs of $G$ is at most 
    $\prod_{i=1}^{\nd}(\card{M_i}+1) \le (\sum_{i=1}^{\nd}(\card{M_i}+1)/\nd)^{\nd}
    \le (n/\nd+1)^{\nd}$.
\end{proof}

Without loss of generality, we select an edge whose endpoints are the minimum indices of vertices in the corresponding module. 
By Proposition \ref{prop:isomorphic}, the algorithm in Section \ref{sec:Exalgo} can be modified to run in time $O^*((n/\nd+1)^{\nd})$.


\begin{theorem}\label{thm:algo:nd}
There is an $O^*((n/\nd+1)^{\nd})$-time algorithm for Arc Kayles.
\end{theorem}
 
The idea can be extended to Colored Arc Kayles and BW-Arc Kayles. 
In $G=(V,\EG\cup \EB\cup \EW)$, two vertices $u,v\in V$ are called \emph{colored twins} if $c(\{u,w\})=c(\{v,w\})$ holds $\forall w \in V\setminus \{u,v\}$. 
We then define the notion of colored neighborhood diversity. 
\begin{definition}\label{def:cnd}
The \emph{colored neighborhood diversity} of $G=(V,E)$ is defined as minimum $\cnd$ such that $V$ can be partitioned into $\cnd$ vertex sets of colored twins.
\end{definition} 
In Colored Arc Kayles or BW-Arc Kayles, we can utilize a partition of $V$ into modules each of which consists of colored twins. 
If we are given a partition of the vertices into colored modules, we can decide the winner of Colored Arc Kayles or BW-Arc Kayles like 
Theorem \ref{thm:algo:nd}. Different from ordinary neighborhood diversity,  it might be hard to compute colored neighborhood diversity in polynomial time. 
\begin{theorem}\label{thm:algo:cnd}
Given a graph $G=(V,\EG\cup \EB\cup \EW)$ with a partition of $V$ into $\cnd$ modules of colored twins, we can compute the winner of Colored Arc Kayles on $G$ in time $O^*((n/\cnd+1)^{\cnd})$.   
\end{theorem}

    

  
     
\begin{figure}[tbp]
 \centering
 \includegraphics[width=0.65\linewidth]{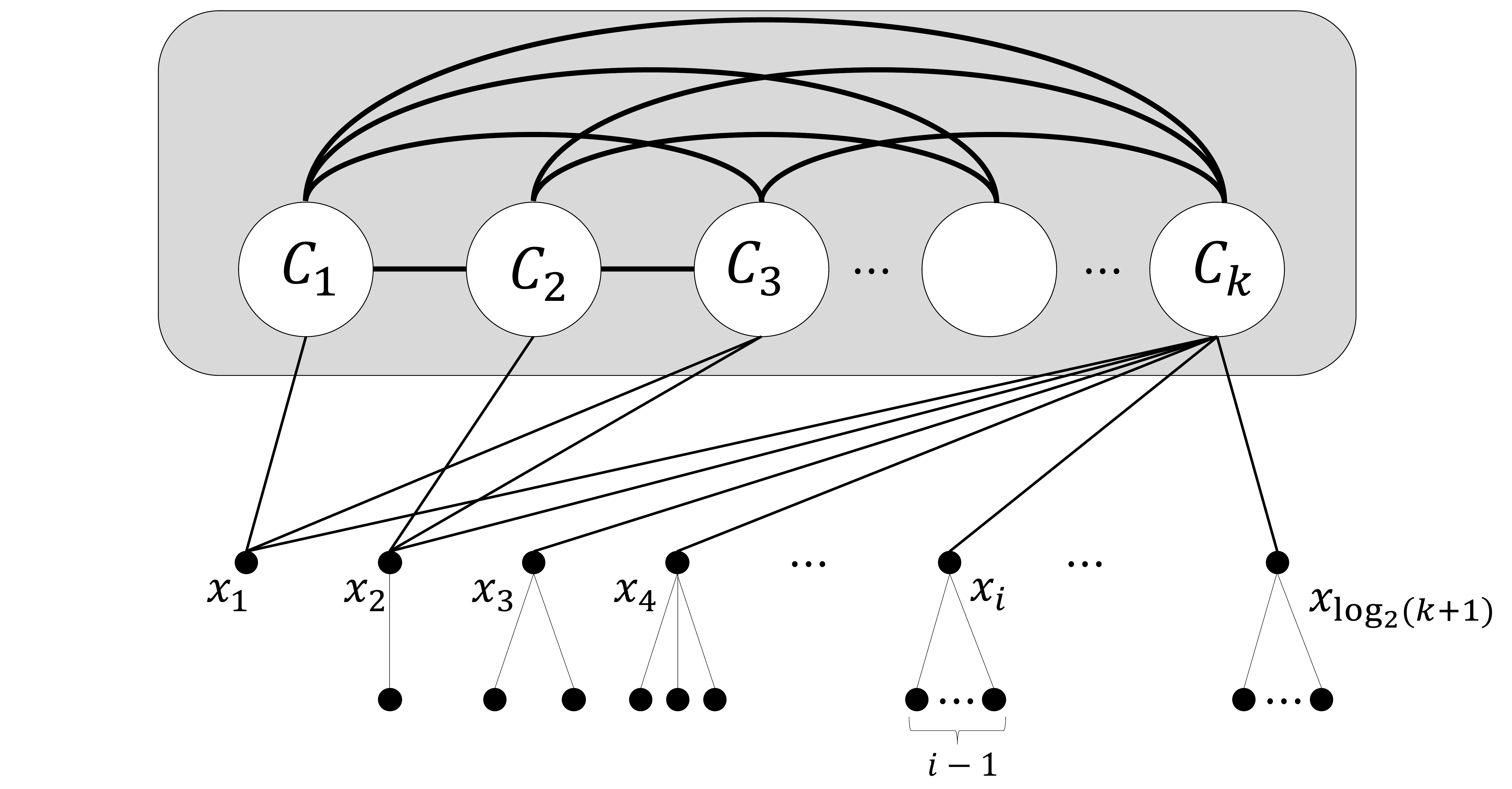}
 \caption{The constructed graph $G$ with neighborhood diversity $\nd=k+2\log_2 (k+1)$.}
 \label{fig:lower:nd}
\end{figure}

In the rest of this section, we give a bad instance for the proposed algorithm as shown in Figure \ref{fig:lower:nd}. 
The result implies that the analysis of Theorem \ref{thm:algo:nd} is 
asymptotically tight. 

\begin{theorem}\label{thm:lower:nd}
There is a graph having at least $(n/\nd+1-o(1))^{\nd(1 - o(1))}$ non-isomorphic positions of Arc Kayles. 
\end{theorem}

\begin{proof}
We construct such a graph $G$. Assume that $k$ is a number forming power of two minus one, that is, $k=2^{k'}-1$. 
First, we prepare $k$ cliques of $s$ vertices, $C_1,\ldots,C_k$, and vertex set $X=\{x_1,x_2,\ldots,x_{\log_2 (k+1)}\}$. The subscript $i$ of $x_i$ represents $i$-th bit used below. 
For each $x_i$, we attach $i-1$ pendant vertices, which is used to distinguish from another $x_j$. For $j$, $\bin(j)$ and $\bin(j,i)$ denote the $j$'s binary representation and its $i$-th bit, respectively. For example, $\bin(6)=110$, 
$\bin(6,1)=0, \bin(6,2)=1, \bin(6,3)=1$, and $\bin(6,i)=0$ for $i\ge 4$. 
We connect the vertices in $C_j$ and vertices in $X$ according to the binary representation of $j$; vertices in $C_j$ are connected with $x_i$ if and only if $\bin(j,i)=1$. Finally, we connect all the vertices in $\bigcup_i C_i$, which form a large clique with size $sk$. 
Figure \ref{fig:lower:nd} shows the constructed graph $G$.

The number of vertices in $G$ is $n=sk + \log_2 (k+1)(\log_2 (k+1)+1)/2$, that is, $s=(n-\log_2 (k+1)(\log_2 (k+1)+1)/2)/k$, 
and the neighborhood diversity of $G$ is $\nd=k+2\log_2 (k+1)$, 
because vertices in each clique are twins, and also pendant vertices connected to each $x_i$ are twins.


We estimate the number of non-isomorphic induced subgraphs of $G$. 
We restrict vertices to delete only from $\bigcup_i C_i$, 
that is, edges to select are inside of $\bigcup_i C_i$. 
Since the number of pendant vertices for each $x_i$ is different, 
$x_i$'s substantially have IDs, and thus vertices from two distinct cliques are distinguishable. Hence, the number of non-isomorphic induced subgraphs obtained by removing edges inside $\bigcup_i C_i$ are decided by the numbers of remaining vertices in $C_i$'s, each of which varies from $0$ to $s$. 
Therefore, the number of non-isomorphic induced subgraphs is at least
\begin{align*}
(s+1)^{k}/2 &= \frac{1}{2}\left(\frac{n-\log_2 (k+1)(\log_2 (k+1)+1)/2}{k}+1\right)^{k}\\
& = \frac{1}{2}\left(\frac{n-o(k)}{k}+1\right)^{k} \ge 
\frac{1}{2}\left(\frac{n}{k}+1-o(1)\right)^{k},
\end{align*}
where the division of $2$ comes from the fact that the number of deleted vertices must be even. 
Since $\nd=k+2\log_2 (k+1)$, $k=\nd - 2\log_2 (k+1)\ge \nd - 2\log_2  \nd$ holds. We thus have lower bound $\left(n/\nd+1-o(1)\right)^{\nd(1 - o(1))}$. 
\end{proof}

\section{Arc Kayles for Trees}\label{sec:trees}
In \cite{BODLAENDER2015165}, Bodlaender et al. show that the winner of Node Kayles on trees can be determined in time $O^*(3^{n/3})=O(1.4423^n)$. 
 It is easy to show by a similar argument that the winner of Arc Kayles can also be determined in time $O(1.4423^n)$. 
 It is also mentioned that the analysis is sharp apart from a polynomial factor because there is a tree for which the algorithm takes $\Omega(3^{n/3})$ time. The example is also available for Arc Kayles; namely, as long as we use the same algorithm, the running time cannot be improved.  
 
 \medskip
 
In this section, we present that the winners of Arc Kayles on trees can be determined in time $O^*(2^{n/2})=O(1.4143^n)$, which is attained by considering a tree (so-called) unordered. Since a similar analysis can be applied to Node Kayles on trees, the winner of Node Kayles on trees can be determined in time $O^*(2^{n/2})$. 

\medskip 

 

Let us consider a tree $T=(V,E)$. 
By Sprague–Grundy theory, if all connected subtrees of $T$ are enumerated, one can determine the winner of Arc Kayles. 
Furthermore, by Proposition \ref{prop:isomorphic}, once a connected subtree $T'$ is listed, we can ignore subtrees isomorphic to $T'$. Here we adopt 
isomorphism of rooted trees. 
\begin{definition}\label{def:isomorphic:tree}
Let $T^{(1)}=(V^{(1)},E^{(1)},r^{(1)})$ and $T^{(2)}=(V^{(2)},E^{(2)},r^{(2)})$ be trees rooted at $r^{(1)}$ and $r^{(2)}$, respectively. Then,  
$T^{(1)}$ and $T^{(2)}$ are called \emph{isomorphic with respect to root} if for any pair of $u,v\in V^{(1)}$ there is a bijection $f: V^{(1)}\to V^{(2)}$ such that $\{u,v\}\in E^{(1)}$ if and only if $\{f(u),f(v)\}\in E^{(2)}$ and $f(r^{(1)})=f(r^{(2)})$. 
\end{definition}
For a tree $T$ rooted at $r$, two subtrees $T'$ and $T''$ are simply said \emph{non-isomorphic} 
if $T'$ with root $r$ and $T''$ with root $r$ are not isomorphic with respect to root. 
Now, we estimate the number of non-isomorphic connected subgraphs of $T$ based on isomorphism of rooted trees. 
For $T=(V,E)$ rooted at $r$, a connected subtree $T'$ rooted at $r$ is called an \emph{AK-rooted subtree} of $T$, if there exists a matching $M  \subseteq E$ such that $T[V\setminus \bigcup M]$ consists of $T'$ and isolated vertices.
Note that $M$ can be empty, AK-rooted subtree $T'$ must contain root $r$ of $T$, and the graph consisting of only vertex $r$ can be an AK-rooted subtree.

\begin{lemma}\label{lem:K-set:tree}
Any tree rooted at $r$ has $O^*(2^{n/2})(=O(1.4143^n))$ non-isomorphic AK-rooted subtrees rooted at $r$.  
\end{lemma}

\begin{proof}
 Let $R(n)$ be the maximum number of non-isomorphic AK-rooted subtrees of any tree rooted at some $r$ with $n$ vertices. 
We claim that $R(n)\le 2^{n/2}-1$ for all $n\geq 4$, which proves the lemma. 

We will prove the claim by induction. For $n\le 4$, the values of $R(n)$'s are as follows: $R(1)=1, R(2)=1, R(3)=2$, and $R(4)=3$. These can be shown by concretely enumerating trees. For example, for $n=2$, a tree $T$ with $2$ vertices is unique, and an AK-rooted subtree of $T$ containing $r$ is also unique, which is $T$ itself. For $n=3$, the candidates of $T$ are shown in Figure \ref{tree1}. For Type A in Figure \ref{tree1}, AK-rooted subtrees are the tree itself and isolated $r$, and for Type B, an AK-rooted subtree is only the tree itself; thus we have $R(3)=2$. Similarly, we can show $R(4)=3$ as seen in Figure \ref{tree2}. 
Note that $R(1)>2^{1/2}-1$, $R(2)=1\le 2^{2/2}-1=1$, $R(3)=2 > 2^{3/2}-1$, and $R(4)=3\le 2^{4/2}-1=3$. This $R(4)$ is used as the base case of induction. 

\begin{figure}[htbp]
\begin{minipage}{0.48\hsize}
 \centering
 \includegraphics[width=0.55\linewidth]{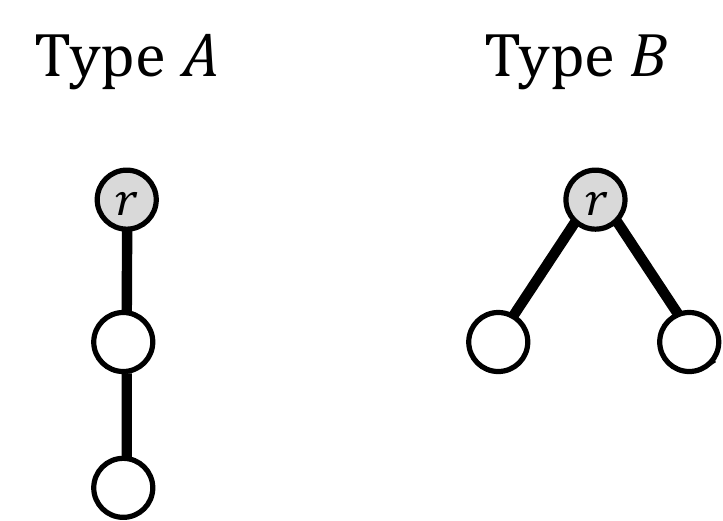}
 \caption{Trees with 3 vertices rooted at $r$}
 \label{tree1}
\end{minipage}
\begin{minipage}{0.48\hsize}
 \centering
 \includegraphics[width=1.05\linewidth]{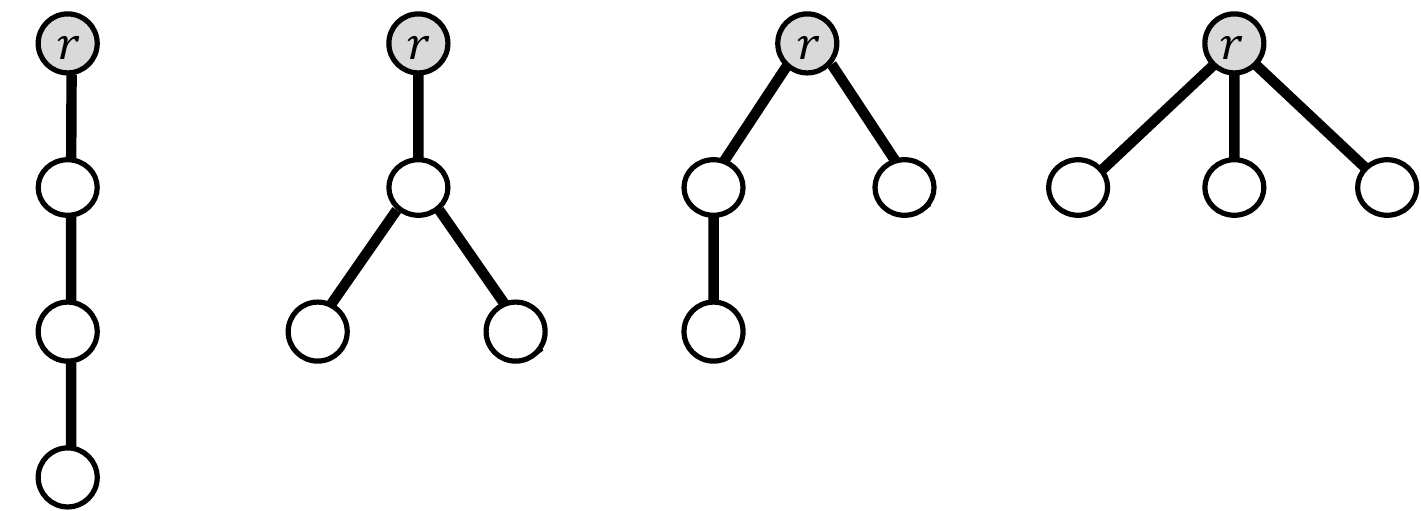}
 \caption{Trees with 4 vertices rooted at $r$}
 \label{tree2}
\end{minipage}
\end{figure}
As the induction hypothesis, let us assume that the claim is true 
for all $n' < n$ except $1$ and $3$, and consider a tree $T$ rooted at $r$ with $n$ vertices. Let $u_1, u_2, \ldots, u_p$ be the children of root $r$, and $T_i$ be the subtree of $T$ rooted at $u_i$ with $n_i$ vertices for $i=1,2,\ldots,p$. 
Note that for an AK-rooted subtree $T'$ of $T$, the intersection of $T'$ and $T_i$ for each $i$ is either empty or an AK-rooted subtree of $T_i$ rooted at $u_i$. Based on this observation, we take a combination of the number of AK-rooted subtrees of $T_i$'s, which gives an upper bound on the number of AK-rooted subtrees of $T$. We consider two cases: (1) for any $i$, $n_i \neq 3$, (2) otherwise. For case (1), the number of AK-rooted subtrees of $T$ is at most 
\[
\prod_{i:n_i>1}(R(n_i)+1)\cdot \prod_{i:n_i=1}1 \le \prod_{i:n_i>1} 2^{n_i/2}=2^{\sum_{i:n_i>1} n_i/2} \le 2^{(n-1)/2} \le 2^{n/2}-1.    
\]
That is, the claim holds in this case. Here, in the left hand of the first inequality, $R(n_i)+1$ represents the choice of AK-rooted subtree of $T_i$ rooted at $u_i$ or empty, 
and ``$1$'' for $i$ with $n_i=1$ represents that $u_i$ needs to be left as is, because otherwise edge $\{r,u_i\}$ must be removed, which violates the condition ``rooted at $r$''.  The first inequality holds since any $n_i$ is not $3$ and thus the induction hypothesis can be applied.  
The last inequality holds by $n\ge 5$.  

For case (2), we further divide into two cases: (2.i), for every $i$ such that $n_i=3$, $T_i$ is Type B, and (2.ii) otherwise. For case (2.i), since an AK-rooted subgraph of $T_i$ of Type B in Figure \ref{tree1} is only $T_i$ itself, the number is $1\le 2^{3/2}-1$. Thus, the similar analysis of Case (1) can be applied as follows: 
\begin{align*}
\prod_{i:n_i\neq 1, 3}(R(n_i)+1) \cdot \prod_{i: T_i \mathrm{{\ is\ Type\ B }}  }(2^{3/2}-1+1) \le \prod_{i:n_i>1} 2^{n_i/2}\le 2^{n/2}-1,     
\end{align*}
that is, the claim holds also in case (2.i). 

Finally, we consider case (2.ii). By the assumption, at least one $T_i$ is Type A in Figure \ref{tree1}. Suppose that $T$ has $q$ children of $r$ forming Type A, which are renamed $T_1,\ldots,T_q$ as canonicalization. Such renaming is allowed because we count non-isomorphic subtrees.  
Furthermore, we can sort AK-rooted subtrees of $T_1,\ldots, T_q$ as canonicalization. Since each Type A tree can form in $T'$ empty, a single vertex, or Type A tree itself, $T_1,\ldots,T_q$ of $T$, the number of possible forms of subforests of $T_1,\ldots,T_q$ of $T$ is 
\[
\multiset{q}{3}=\binom{q+2}{2}. 
\]
Since the number of subforests of $T_i$'s other than $T_1,\ldots,T_q$ are similar evaluated as above, we can bound the number of AK-rooted subtrees by \[
\binom{q+2}{2} \prod_{i:i>q} 2^{n_i/2} \le \frac{(q+2)(q+1)}{2} 2^{\sum_{{i:i>q}} n_i/2} \le \frac{(q+2)(q+1)}{2} 2^{(n-3q-1)/2}.  
\]
Thus, to prove the claim, it is sufficient to show that $(q+2)(q+1)2^{(n-3q-3)/2}\le 2^{n/2}-1$ for any pair of integers $n$ and $q$ satisfying $n\ge 5$ and $1\le q\le (n-1)/3$.  
This inequality is transformed to the following
\[
\frac{(q+1)(q+2)}{2^{\frac{3(q+1)}{2}}} \le 1-\frac{1}{2^{\frac{n}{2}}}.
\]
Since the left hand and right hand of the inequality are monotonically decreasing with respect to $q$ and monotonically increasing with respect to $n$, respectively, the inequality always holds if it is true for $n=5$ and $q=1$. In fact, we have 
\[
\frac{(1+1)(1+2)}{2^{\frac{3(1+1)}{2}}}= \frac{3}{4}=1 - \frac{1}{2^2} \le 1-\frac{1}{2^{\frac{5}{2}}}, 
\]
which completes the proof. 
\end{proof}

\begin{theorem}
The winner of Arc Kayles on a tree with $n$ vertices can be determined in time $O^*(2^{n/2})=O(1.4143^n)$.
\end{theorem}

\bigskip

In the rest of this section, we mention that we can determine the winner of Node Kayles for a tree in the same running time as Arc Kayles. The outline of the proof is also almost the same as Arc Kayles. Only the difference is to utilize the notion of \emph{NK-rooted subtree} instead of 
AK-rooted subtree for Arc Kayles. 
For $T=(V,E)$ rooted at $r$, a connected subtree $T'$ rooted at $r$ is called an NK-rooted subtree of $T$, if there exists an independent set $U  \subseteq V$ such that $T[V\setminus N[U]]=T'$. 


\begin{lemma}
Any tree rooted at $r$ has $O^*(2^{n/2})(=O(1.4143^n))$ non-isomorphic NK-rooted subtrees rooted at $r$.  
\end{lemma}

\begin{proof}
Let $\hat{R}(n)$ be the maximum number of non-isomorphic NK-rooted subtrees of any tree rooted at some $r$ with $n$ vertices. 
Similarly to Arc Kayles, we can show that $\hat{R}(n)\le 2^{n/2}-1$ for all $n\geq 4$ by induction. For $n\le 4$, it is easy to see that the values of $\hat{R}(n)$'s coincide with those of $R(n)$'s: $\hat{R}(1)=R(1)=1, \hat{R}(2)=R(2)=1, \hat{R}(3)=R(3)=2$ (see Figure \ref{tree1}), and $\hat{R}(4)=R(4)=3$ (see Figure \ref{tree2}). That is, $\hat{R}(n)\le 2^{n/2}-1$ does not hold for $n=1$ and $3$, whereas it holds for $n=2$ and $4$, which shows the base step of the induction. We then consider the induction step. 

In the induction step, we again take the same strategy as Arc Kayles; 
we take a combination of the number of NK-rooted subtrees of $T_i$'s, which gives an upper bound on the number of NK-rooted subtrees of $T$. Since all the arguments use the same induction hypothesis and the same values of $\hat{R}(n)=R(n)$ for $n=1,2,3,4$, the derived bound is the same as Arc Kayles. 
\end{proof}

\subsection*{Acknowledgments}
This work is partially supported by JSPS KAKENHI JP20H05967, JP21H05852, JP21K17707, JP21K19765, JP21K21283, JP22H00513.

\bibliographystyle{plain}
\bibliography{sankou}
\end{document}